\documentclass[submission,copyright]{eptcs}
\providecommand{\event}{NCMA 2024}

\usepackage{amsmath}
\usepackage{amssymb}
\usepackage{tabularray}
\usepackage{subcaption}
\usepackage{algorithm}
\usepackage{algpseudocode}
\usepackage{enumitem}
\setlist{noitemsep, nolistsep}

\title{A GLR-like Parsing Algorithm for Three-Valued Interpretations of Boolean Grammars with Strong Negation\thanks{Supported by the University of Debrecen Scientific Research Bridging Fund (DETKA).}}
\author{Patrik Adrián
\institute{Faculty of Informatics, University of Debrecen,\\
Kassai út 26, 4028 Debrecen, Hungary}
\email{adrianpatrik@mailbox.unideb.hu}
\and
György Vaszil
\institute{Faculty of Informatics, University of Debrecen,\\
Kassai út 26, 4028 Debrecen, Hungary}
\email{vaszil.gyorgy@inf.unideb.hu}
}

\def\titlerunning{A GLR-like algorithm for three-valued interpretations of Boolean grammars}
\def\authorrunning{P. Adrián \& Gy. Vaszil}

    \algtext*{EndIf}%
    \algtext*{EndFor}%
    \algtext*{EndWhile}%
\begin{document}
\maketitle

\begin{abstract}
Boolean grammars generalize context-free rewriting by extending the possibilities when dealing with different rules for the same nonterminal symbol. By allowing not only disjunction (as in the case of usual context-free grammars), but also conjunction and negation as possible connections between different rules with the same left-hand side,  they are able to simplify the description of context-free languages and characterize languages that are not context-free. The use of negation, however, leads to the possibility of introducing rules that interplay in such a way which is problematic to handle in the classical, two-valued logical setting. Here we define a three valued interpretation  to deal with such contradictory grammars using a method introduced originally in the context of logic programming, and present an algorithm to determine the membership status of strings with respect to the resulting three valued languages.
\end{abstract}

Ever since their publication in 1956, context-free grammars (CFG) of Chomsky \cite{Chomsky1956} have served as the ubiquitous tool for formal grammar specification, thanks to their easy-to-understand semantics and admission of simple parsing algorithms. Other formalisms, such as tree adjunct grammars \cite{Joshi1975}, parsing expression grammars \cite{Ford2004} and others have since been developed, partially to address the inadequacy of CFGs to fully describe natural languages, but none have been as successful as CFGs themselves.

Even though the original semantics of a CFG are defined in terms of a rewrite system over an alphabet of terminal and nonterminal symbols, the parsing problem may just as well be seen as a problem of logic, where grammar rules serve as rules of inference and parsing is a search for the proof of the root proposition. The Boolean grammars of Okhotin \cite{Okhotin2004} build upon this interpretation and extend traditional CFGs with conjunction (intersection) and negation (complementation) operations. When interpreted over a two-valued, classical logic, a Boolean grammar may be contradictory and not have a satisfying solution. In a three-valued logic where indeterminacy is a truth value that is stable under negation, such contradictions become tractable.

Our work involves the development of a parser based on a generalized LR (GLR) method for the entailment semantics of a Boolean grammar. Okhotin's GLR-like parser \cite{Okhotin2006} for Boolean grammars works on a two-valued foundation and is not general in the sense of a generalized parser, as it cannot handle certain classes of grammars. His algorithm realizes negation by the deletion of edges from the graph structured stack (GSS) used by the algorithm; our solution is more in the spirit of the original GLR (see \cite{Tomita1985a}), which uses a monotone approach to parsing, where edges are only created, never removed.

We were motivated by the work of Kountouriotis et al. \cite{Kountouriotis2009} that described a tabular parser for the well-founded semantics of a Boolean grammar. The well-founded semantics, originally introduced in \cite{Gelder1991} is a three-valued semantic interpretation of a logic program that builds on a restricted version of the closed world semantics and a closely associated rule of inference often referred to as ``negation as failure''. The (also three-valued) semantic model of \cite{Fitting1985} builds on what is very close to the open world semantics and infers knowledge based only on entailment, rather than failure to be proven true.

The reader would rightfully expect the toy grammar with the singular rule $S \to \neg S$ to be self-contradictory and not have any two-valued models. In a three-valued setting, the language defined by such a grammar has an indeterminate relation to all strings of the underlying alphabet, i.e. it neither contains, nor excludes them. A more interesting case is the similar grammar with the rule $S \to S$, which, unlike the previous example, does have a two-valued model; in fact, every conceivable language models this grammar. The well-founded model of this grammar is the language that excludes all strings. One might, however, argue that the choice made here is rather arbitrary and is only a leftover from the two-valued world; the ``correct'' three-valued solution here is that one also cannot determine the containment status of words within this language; this time not because of inconsistency, but inadequacy.
The Fitting-semantics of logic programs (and by extension, Boolean grammars) is based on the latter philosophy, and considers the containment status to be determinate if and only if it cannot be otherwise (i.e. it is entailed by the axioms, here implied by the rules of the grammar).

\section{Preliminaries}

We base our discussion on a highly restricted fragment of first order logic that, for the lack of function symbols, variables and quantifiers, we consider to be effectively propositional.

An atom is of the form $P(c)$, where $c$ is a constant symbol and $P$ is a unary predicate. A formula is either an atom or is built using the usual connectives $\lnot $, $\land $ and $\lor $, in decreasing order of precedence. The set of all constant symbols is the Herbrand-universe ($\mathcal{U}$) and the set of atoms are the Herbrand-base ($\mathcal{B}$) of the language.

A rule is of the form $A \leftarrow \phi $ where $A$ is an atom and $\phi $ is a formula. The symbol $A$ is the head, $\phi $ is the body of the rule. A set of rules is well-formed if and only if (iff) no two rules have the same head and, for every atom $A'$ that appears anywhere within the body of a rule, a rule with head $A'$ exists, i.e. we require that each atom is defined exactly once. The reader may assume that we are only dealing with well-formed rulesets.

We consider the set of possible truth values $\mathbb{B} = \{\top , \bot , \backsim \}$ representing truth, falsity and a third judgement understood as being indeterminate. The usual Boolean operations are as in Figure \ref{fig:kleene-strong-3vl}, also known as Kleene's strong three-valued logic. A valuation is a function $\mathcal{B}\to \mathbb{B}$ that assigns a truth value to every atom in the Herbrand-base. We define the strict partial order $\prec $ over $\mathbb{B}$ as $\backsim  \prec  \bot $ and $\backsim  \prec  \top $, with $\bot $ and $\top $ unrelated. The relation $\preceq $ is the reflexive closure of $\prec $. A valuation $I_1$ is no more certain than $I_2$, written as $I_1 \preceq  I_2$ iff for all atoms $A$ it holds that $I_1(A) \preceq  I_2(A)$, i.e. $I_2$ changes at most the truth values of atoms that are indeterminate in $I_1$. For an arbitrary formula $\phi $, it holds that $I_1(\phi ) \preceq  I_2(\phi )$.

\begin{figure}[ht]
    \centering
    \begin{subfigure}{.20\textwidth}
    \centering
    \begin{tabular}{|c||c|}%{
%        hlines = {1pt},
%        vlines = {1pt},
%        hline{2} = {2pt},
%        vline{2} = {2pt},
%        cells = {mode=math,halign=c},
%        rows={ht=\baselineskip},
%        stretch=0,
%    }
\hline
        $\neg$ & \space \\ \hline\hline
        $\top$ & $\bot$ \\ \hline
        $\backsim$ & $\backsim$ \\ \hline
        $\bot$ & $\top$ \\ \hline
    \end{tabular}
    \caption{Negation.}
    \end{subfigure}
%    \hfill
    \begin{subfigure}{.30\textwidth}
    \centering
    \begin{tabular}{|c||c|c|c|}%{
%        hlines = {1pt},
%        vlines = {1pt},
%        hline{2} = {2pt},
%        vline{2} = {2pt},
%        cells = {mode=math,halign=c},
%        rows={ht=\baselineskip},
%        stretch=0,
%    }
\hline
        $\land$ & $\top$ & $\backsim$ & $\bot$ \\ \hline\hline
        $\top$ & $\top$ & $\backsim$ & $\bot$ \\ \hline
        $\backsim$ & $\backsim$ & $\backsim$ & $\bot$ \\ \hline
        $\bot$ & $\bot$ & $\bot$ & $\bot$ \\ \hline
    \end{tabular}
    \caption{Conjunction.}
    \end{subfigure}
%    \hfill
    \begin{subfigure}{.30\textwidth}
    \centering
    \begin{tabular}{|c||c|c|c|}%{
%        hlines = {1pt},
%        vlines = {1pt},
%        hline{2} = {2pt},
%        vline{2} = {2pt},
%        cells = {mode=math,halign=c},
%        rows={ht=\baselineskip},
%        stretch=0,
%    }
\hline
        $\lor$ & $\top$ & $\backsim$ & $\bot$ \\ \hline\hline
        $\top$ & $\top$ & $\top$ & $\top$ \\ \hline
        $\backsim$ & $\top$ & $\backsim$ & $\backsim$ \\ \hline
        $\bot$ & $\top$ & $\backsim$ & $\bot$ \\ \hline
    \end{tabular}
    \caption{Disjunction.}
    \end{subfigure}
    \hfill

    \caption{Kleene's strong three-valued connectives.}
    \label{fig:kleene-strong-3vl}
\end{figure}

A rule $A \leftarrow \phi $ is satisfied by the valuation $I$ iff $I(A) = I(\phi )$, i.e. the truth value of its left-hand side is the same as the value of the formula on its right-hand side, when evaluated over $I$. A set of rules $\Pi $ is satisfied by $I$ iff all rules in $\Pi $ are satisfied by $I$. A set of ground rules $\Pi $ may be written as a (potentially countably infinite) vector equation $\mathbf{A} \equiv \boldsymbol{\phi}$ where $\mathbf{A}$ contains atoms and $\boldsymbol{\phi}$ contains formulas. A valuation $I$ is the solution of this vector equation iff $I(\mathbf{A}) = I(\boldsymbol{\phi})$, where the elements are evaluated memberwise.

We will now describe deduction based on the semantics defined by Fitting \cite{Fitting1985}.

Given a set of rules $\Pi $, the operator $\phi $ maps an arbitrary valuation $I$ to its $\phi $-successor $\phi (I)$ such that $$\phi (I)(\mathbf{A})=I(\boldsymbol{\phi}).$$

Let $I_\backsim $ be the null valuation such that $I_\backsim (A) = \mathord \backsim $ for all atoms $A$. Let $n$ be a finite ordinal and $\omega _0$ be the first infinite ordinal. We define
\begin{align*}
\boldsymbol{I}_0 &= I_\backsim  \\
\boldsymbol{I}_n &= \phi (\boldsymbol{I}_{n-1}) \\
\boldsymbol{I}_{\omega _0} &= \sup_{n<\omega _0} \boldsymbol{I}_n
\end{align*}
where the supremum is taken over $\preceq $ and is equal to $\bigcup_{n < \omega _0} \boldsymbol{I}_n$ where union is understood as $(I_1 \cup I_2)(A) = \max_{\preceq} \{ I_1(A), I_2(A) \}$. The sequence $\boldsymbol{I}$ is monotone in $\preceq $ and has a supremum $\omega $ that we call the entailment model of $\Pi $.

An important property of $\boldsymbol{I}$ is that it is monotone with regards to $\preceq $, i.e. it never ``retracts'' any conclusion already made. Since $\omega  = \bigcup_{n < \omega _0} \boldsymbol{I}_n$, any atom that has an assigned truth value in $\omega $ must have one in $\boldsymbol{I}_n$ for some finite $n$.

In the original setting of logic programming, where arbitrary terms of first-order logic may be formed, determinacy is only semidecidable, though our formulas will be constructed such that it is fully decidable. This is due to the dependency set (the transitive closure of the set of atoms occurring in $\phi$), for any rule $A \gets \phi$, being finite for every atom, therefore an evaluation procedure requiring only a finite number of evaluations to determine the status of any $A$.

\section{Three-valued languages and semantics}

Given an alphabet $\Sigma $, $\Sigma ^\ell$ is the set of all strings (words) of length $\ell$ and $\Sigma ^*$ is $\bigcup_{i \geq 0} \Sigma ^i$. A (classical) language over $\Sigma $ is a (possibly improper) subset of $\Sigma ^*$.

An $n$-partition of a word $w$ is the tuple of words $w_1, \ldots, w_n$ such that $w = w_1 \cdots w_n$, where $w_1 \cdots w_n$ is the concatenation of $w_1, \ldots, w_n$. Similarly, for natural numbers, an $n$-partition of a natural number $\ell$ is an element of $\mathbb{N}^n$ whose members add up to $\ell$. We will take advantage of the natural isomorphism between the partitions of a natural number $\ell$ and those of a word $w$ with $|w| = \ell$.

The concatenation of languages $L_1, \ldots, L_n$, denoted as $L_1 \cdots L_n$ is the language of words $w$ such that there exists a partition $w = w_1 \cdots w_n$ such that $w_i \in L_i$, for all $1 \leq i \leq n$.

A three-valued language is a pair of languages $L = \langle L^\top , L^\bot  \rangle$ over an alphabet $\Sigma $ such that $L^\top  \cap  L^\bot $ is empty. Notice that it is \emph{not} required that $L^\top  \cup  L^\bot  = \Sigma ^*$. We define the following operations on three-valued languages (we use $L | {}^\ell$ to denote the set of words in $L$ that are exactly of length $\ell$):
\begin{align*}
\overline{L}   &= \langle L^\bot , L^\top  \rangle \\
L_1 \cup  \cdots \cup  L_n &= \langle L_1^\top  \cup  \cdots \cup  L_n^\top , L_1^\bot  \cap  \cdots \cap  L_n^\bot  \rangle \\
L_1 \cap  \cdots \cap  L_n &= \langle L_1^\top  \cap  \cdots \cap  L_n^\top , L_1^\bot  \cup  \cdots \cup  L_n^\bot  \rangle \\
L_1 \cdots L_n &= \left\langle \bigcup_{(l\geq 0)} \bigcup_{(p_1 + \cdots + p_n = l)} \bigcap_{(i\leq n)} \Sigma ^{p_1} \cdots \Sigma ^{p_{i-1}} (L_i^\top  | ^{p_i}) \Sigma ^{p_{i+1}} \cdots \Sigma ^{p_n}\right.,
\\
& \hspace*{4.5cm} \left.\bigcup_{(l\geq 0)} \bigcap_{(p_1 + \cdots + p_n = l)} \bigcup_{(i\leq n)} \Sigma ^{p_1} \cdots \Sigma ^{p_{i-1}} (L_i^\bot  |^{p_i}) \Sigma ^{p_{i+1}} \cdots \Sigma ^{p_n} \right\rangle
\end{align*}
These definitions agree with those in \cite{Kountouriotis2009}, in particular
\begin{itemize}
\item
a word is an element of ${(L_1\cdots L_n)}^\top$ iff it has an $n$-partition such that for all $1\leq i\leq n$ the $i$th part belongs to $L_i$ and
\item
a word is an element of ${(L_1\cdots L_n)}^\bot$ iff in every $n$-partition there exists an $1\leq i\leq n$ such that the $i$th part is excluded from $L_i$.
\end{itemize}

The characteristic function of a three-valued language $L = \langle L^\top , L^\bot  \rangle$ is the function $L : \Sigma ^* \to  \mathbb{B}$ such that $$L(w) = \begin{cases} \top & \text{if } w \in L^\top , \\ \bot & \text{if } w \in L^\bot , \\ \backsim & \text{otherwise.} \end{cases}$$ We will write $w \in L$ for $w \in L^\top $ and $w \not\in L$ for $w \in L^\bot $; note that containment is not dichotomous. The characteristic functions of the above are
\begin{align*}
\overline{L}(w) &= \lnot L(w) \\
(L_1 \cup  \cdots \cup  L_n)(w) &= \bigvee_{i\leq n} L_i(w) \\
(L_1 \cap  \cdots \cap  L_n)(w) &= \bigwedge_{i\leq n} L_i(w) \\
(L_1 \cdots L_n)(w) &= \bigvee_{w = w_1 \cdots w_n}\bigwedge_{i\leq n}L_i(w_i)
\end{align*}

A three-valued language may either include, exclude any given string or the containment may be indeterminate. Indeterminacy may, informally, be understood as a sort of ``weak exclusion'' that is unsuitable for further deduction. The set of all three-valued languages over the alphabet $\Sigma $ will be denoted by $\mathcal{L}$.

\subsection{Boolean grammars}

A Boolean grammar is a triple $G = \langle \mathbf{V}, \Sigma , \mathbf{P} \rangle$ where $\mathbf{V}$ is the a of grammar variables (nonterminals), $\Sigma $ is an alphabet (terminals) and $\mathbf{P}$ is a set of grammar rules (productions). We will use $\Gamma = \mathbf{V} \cup \Sigma  \cup \{ \epsilon \}$ to denote the complete set of grammar symbols ($\epsilon \not\in \mathbf{V} \cup \Sigma $).

We define grammar expressions and grammar rules as follows.
\begin{itemize}
    \item Members of $\Gamma$ are grammar expressions.
    \item If $\phi$ is an expression, then $\neg\phi$ is a negated expression.
    \item If $\phi_1, \ldots, \phi_n$ are expressions, then $\phi_1 \vee \cdots \vee \phi_n$ is a disjunctive expression.
    \item If $\phi_1, \ldots, \phi_n$ are expressions, then $\phi_1 \wedge \cdots \wedge \phi_n$ is a conjunctive expression.
    \item If $\phi_1, \ldots, \phi_n$ are expressions, then $\phi_1 \cdots \phi_n$ is a concatenation expression.
    \item If $\phi$ is an expression and $X \in \mathbf{V}$, then $X \to \phi$ is a grammar rule and $X$ is its head.
\end{itemize}

Rules are the top-level constructs of a Boolean grammar and are not expressions themselves. A Boolean grammar is well-formed if, for all $X \in \mathbf{V}$, there is exactly one rule whose head is $X$. Furthermore, we assume that $n > 1$ and that no direct subexpression of a grammar expression is of the same kind as its parent.

Given a set of grammar variables $\mathbf{V}$, an interpretation is a function $I: \mathbf{V} \to  \mathcal{L}$. We may naturally extend it to arbitrary expressions as follows:
\begin{itemize}
    \item $I(\epsilon) = \langle \emptyset, \overline\emptyset \rangle$;
    \item $I(t) = \langle \{ t \}, \overline{\{ t \}} \rangle$ where $t \in \Sigma $;
    \item $I(\neg \phi) = \overline{I(\phi)}$;
    \item $I(\phi_1 \vee \cdots \vee \phi_n) = I(\phi_1) \cup \cdots \cup I(\phi_n)$;
    \item $I(\phi_1 \wedge \cdots \wedge \phi_n) = I(\phi_1) \cap \cdots \cap I(\phi_n)$;
    \item $I(\phi_1 \cdots \phi_n) = I(\phi_1) \cdots I(\phi_n)$.
\end{itemize}
All complements are understood with respect to a universe of $\Sigma ^*$.

A grammar rule $X \to \phi$ is to be understood as an equation $I(X) = I(\phi)$. An interpretation $I$ is a model (a solution) of a grammar if and only if all grammar rules hold in $I$.

Somewhat similar in spirit to the naturally reachable semantics of Okhotin \cite{Okhotin2004}, the entailment semantics of Boolean grammars may be defined using an iterative approach. Let $\langle X_1, \ldots, X_{|\mathbf{V}|} \rangle$ be a particular ordering of the grammar variables. We may then write an interpretation $I$ as a vector of languages $\langle I(X_1), \ldots, I(X_{|\mathbf{V}|}) \rangle$. Starting from $\boldsymbol{I}_0 = \langle \langle \emptyset, \emptyset \rangle, \ldots, \langle \emptyset, \emptyset \rangle \rangle$ as the null interpretation, we may assign the next interpretation $\boldsymbol{I}_{n+1}$ as $\langle \boldsymbol{I}_n(\phi_1), \ldots, \boldsymbol{I}_n(\phi_{|\mathbf{V}|}) \rangle$, where $\phi_i$ is the definition of variable $X_i$, i.e. there is a rule $X_i \to \phi_i$ in the grammar. The sequence always converges (in at most a countably infinite number of steps) and provides a natural foundation of what we consider to be a natural three-valued semantics of a Boolean grammar. The convergence also holds if only one element is updated at a time, i.e. if $i$ is arbitrarily chosen between $1$ and $|\mathbf{V}|$ (assuming that each value is eventually picked a sufficient number of times), then $\boldsymbol{I}_{n+1} = \langle \boldsymbol{I}_n(X_1), \ldots, \boldsymbol{I}_n(\phi_i) \ldots, \boldsymbol{I}_n(X_{|\mathbf{V}|}) \rangle$.

This a construction, while it serves as a natural semantic model for a Boolean grammar, is not particularly useful for parsing. The following approach ultimately defines the same model but does so for a single word at a time, using a particular construction of logic rules based on the characteristic functions. This is the theoretical foundation of how our parser makes inferences.

Let the (countably infinitely many) constants of our language of logic be the words of $\Sigma^*$ and the (unary) predicate symbols be members of $\Gamma$. We will construct a countably infinite set of logic rules, one for each word and grammar rule, that expresses their semantics.

We define a function $\varrho(\phi, w)$ that takes an arbitrary grammar expression $\phi$ and a variable $w$ in the language of logic and maps it to an open (parametric) logic formula as follows:
\begin{itemize}
    \item if $\phi \in \Gamma$, then $\varrho(\phi, w)$ is $\phi(w)$;
    \item if $\phi$ is $\neg \psi$, then $\varrho(\phi, w)$ is $\neg\varrho(w)$;
    \item if $\phi$ is $\psi_1 \vee \cdots \vee \psi_n$, then $\varrho(\phi, w)$ is $(\varrho(\psi_1, w) \vee \cdots \vee \varrho(\psi_1, w))$;
    \item if $\phi$ is $\psi_1 \wedge \cdots \wedge \psi_n$, then $\varrho(\phi, w)$ is $(\varrho(\psi_1, w) \wedge \cdots \wedge \varrho(\psi_1, w))$;
    \item if $\phi$ is $\psi_1 \cdots \psi_n$, then $\varrho(\phi, w)$ is $\bigvee_{w = w_1 \cdots w_n} \bigwedge_{i \leq n} \varrho(\psi_n, w_n)$.
\end{itemize}
where the variables $w_1, \ldots, w_n$ are new variables. The parametric forms of the rules are:
\begin{itemize}
    \item for each grammar rule $X \to \phi$ we generate $X(w) \gets \varrho(\phi, w)$;
    \item for each terminal $t \in \Sigma  \cup \{ \epsilon \}$ we generate $t(w) \gets t = w$.
\end{itemize}

As a final step, the variables are substituted by the constants, i.e. the words over $\Sigma$. This results in a total of $|\mathbf{P}| + |\Sigma | + 1$ rules for each word in $\Sigma^*$.

These rules together define the intended meaning of a Boolean grammar. Note that even though the set of rules is infinite (as there are infinitely many words in $\Sigma^*$), the value of every atom is defined, both directly and indirectly, through others with a word length that is not greater than itself. In the worst case, the number of atoms that need to be evaluated to determine the valuation of a string is the number of distinct substrings times the number of symbols, i.e. $(1 + \frac{1}{2} \cdot |w| \cdot ({|w|} + 1)) \cdot |\Gamma|$, which is quadratic in the length of the string.

We shall illustrate the above with the example grammar taken from \cite{Kountouriotis2009} whose language $S$ includes precisely the strings that are of the form $ww$ over an alphabet $\{ a, b \}$:
\begin{align*}
    S &\to \lnot (A \lor B \lor AB \lor BA) \\
    A &\to CAC \lor a \\
    B &\to CBC \lor b \\
    C &\to a \lor b
\end{align*}

The open (parametric) rules generated for the above grammar are:
\begin{align*}
    \epsilon(w) \gets {}& w = \epsilon\\
    a(w) \gets {}&w = a \\
    b(w) \gets {}&w = b \\
    S(w) \gets {}&\lnot \left(A(w) \lor B(w) \lor \bigvee_{w = w\sb{1} w\sb{2}} \left[ A(w\sb{1}) \land B(w\sb{2}) \right] \lor \bigvee_{w = w\sb{1}w\sb{2}} \left[ B(w\sb{1}) \land A(w\sb{2}) \right]\right) \\
    A(w) \gets {}&\bigvee_{w = w\sb{1}w\sb{2}w\sb{3}} \left[ C(w\sb{1}) \land A(w\sb{2}) \land C(w\sb{3}) \right] \lor a(w) \\
    B(w) \gets {}&\bigvee_{w = w\sb{1}w\sb{2}w\sb{3}} \left[ C(w\sb{1}) \land B(w\sb{2}) \land C(w\sb{3}) \right] \lor b(w) \\
    C(w) \gets {}&a(w) \lor b(w)
\end{align*}

Finally, we substitute words into the parameter $w$. As the instantiated set is infinite, we will only demonstrate some rules using $abab$ and its substrings.

\begin{align*}
    \allowdisplaybreaks
    S(abab) \gets {}& \lnot \Big(A({abab}) \lor B({abab}) \lor {}\\
        &\ \ \ \ \big(A(\epsilon) \land B({abab}) \lor
         A(a) \land B(bab) \lor
         A(ab) \land B(ab) \lor {}
         \\
        &\ \ \ \ \ A(aba) \land B(b) \lor
         A(abab) \land B(\epsilon)\big) \lor {}
         \\
        &\ \ \ \ \big(B({\epsilon}) \land A({abab}) \lor
         B(a) \land A(bab) \lor
         B(ab) \land A(ab) \lor {}
         \\
        &\ \ \ \ \ B(aba) \land A(b) \lor
         B(abab) \land A({\epsilon})\big)\Big) \\
        \vdots & \\
    A(\epsilon) \gets {}& \big(C(\epsilon) \land A(\epsilon) \land C(\epsilon)\big) \lor a(\epsilon)
    \\
    A(a)  \gets {}&
        \big(C(a) \land A(\epsilon) \land C(\epsilon) \lor
        C(\epsilon) \land A(a) \land C(\epsilon) \lor
        C(\epsilon) \land A(\epsilon) \land C(a)\big)
        \lor a(a)
        \\
        \vdots & \\
    A(aba) \gets {}&
        \big(C(\epsilon) \land A(\epsilon) \land C(aba) {} \lor
        C(\epsilon) \land A(a) \land C(ba) \lor
        C(\epsilon) \land A(ab) \land C(a) \lor {}
        \\
        &\  C(\epsilon) \land A(aba) \land C(\epsilon) \lor
        C(a) \land A(\epsilon) \land C(ba) \lor
        C(a) \land A(b) \land C(a) \lor {}
        \\
        &\  C(a) \land A(ba) \land C(\epsilon) \lor
        C(ab) \land A(\epsilon) \land C(a) \lor
        C(ab) \land A(a) \land C(\epsilon) \lor {}
        \\
        &\ C(aba) \land A(\epsilon) \land C(\epsilon)\big) \lor
        a(aba) \\
        \vdots & \\
    C(a) \gets {}& a(a) \lor b(a) \\
    C(ab) \gets {}& a(ab) \lor b(ab) \\
    \vdots & \\
    a(a) \gets {}& a = a \\
    a(b) \gets {}& b = a \\
    \vdots &
\end{align*}

The interested reader may want to determine the status of the words $a$, $b$ and $ab$ in $S$ of the grammar
\begin{align*}
    A &\to \epsilon \lor A \\
    S &\to A b
\end{align*}
over the alphabet $\{ a, b \}$\footnote{Excluded, included and indeterminate.}.

\section{The Boolean GLR parser}

First we give a very short review of the GLR algorithm and some of its modifications we build our variant upon.

The LR automaton is essentially a Rabin-Scott construction of a trivial nondeterministic pushdown automaton for a context-free grammar. Whenever a (context-free) rule $A \to \alpha \bullet  B \beta$ is being read, with the dot signaling the current position up to which it has already been recognized, it is allowed to transition to any rule $B \to \gamma$ without the consumption of any input. Transitions not consuming any input are called $\epsilon$-transitions and their closure forms the states of the LR automaton. The LR parser is a deterministic simulation of this automaton using a single stack, and has two main operations, \emph{shift} and \emph{reduce}, roughly equivalent to the stack operations \emph{push} and \emph{pop}. Shifting happens when the automaton reads input and pushes the new state on the stack; reduction consists of the removal of as many states as there are on the right-hand side of a rule and a new state, corresponding to having read the left-hand side of the rule, is pushed in their place. The automaton has a special state that signals the recognition of the start symbol and serves as a terminator.

Given that some context-free grammars are not deterministically recognizable using the LR algorithm, as the parsing actions are ambiguous (shift/reduce or reduce/reduce conflict), the first attempts to broaden the algorithm's applicability involved the use of lookaheads to assist the decision process. Though an improvement over the naïve design, lookaheads only generalize LR parsing to deterministic context-free grammars, which is a proper subset of all CFGs.

Viewing the stack as a linear directed acyclic graph (DAG), it is possible to efficiently simulate nondeterminism by generalization of the ``graph-stack'' into a nonlinear DAG, exploring all paths the LR automaton might take . The resulting structure is often termed a graph structured stack (GSS) and can be seen as a generalization of a stack, where every path ending at the root is a record of a possible stack of the LR automaton. Note that it is possible, but not necessary, to use lookaheads to disambiguate actions of the GLR algorithm.

The original GLR has a weakness in design when it comes to nullable rules (in a CFG, a rule is nullable iff every symbol on its right-hand side is nullable, i.e. derives the empty string), namely that edges corresponding to nulled deductions are still created in the GSS. This not only negatively affects the algorithm's efficiency, but also raises problems of correctness on a general CFG when certain rules with nullable right-ends are concerned.

One solution to the problem is the $\epsilon$-GLR construction of Nederhof and Sarbo \cite{Nederhof1996} that modifies the Rabin-Scott closure so that the closure of an item $A \to \alpha \bullet  B \beta$ not only includes items of $B \to \bullet  \gamma$, but -- iff $B$ is nullable -- also $A \to \alpha B \bullet  \beta$. This modification prevents the creation of nulled edges in the graph at the cost of more complicated reductions, as now edges corresponding to nullable symbols might or might not be absent from the GSS. Another approach is the RNGLR of Scott and Johnstone \cite{scott_right_2006}, which performs reductions early when all symbols to the right of the dot are nullable.

These algorithms may not be cubic in the worst case, as the path scanning (determining the GSS nodes at which a reduction may end) may be of complexity $O({|w|}^{n - 1})$ for a path with $n$ components, for a total runtime of $O({|w|}^{n + 1})$. One may rewrite the grammar in Chomsky Normal Form to guarantee cubic runtime, which may, depending on the implementation and applied postprocessing of the parsing results, completely destroy its semantic structure. The BRNGLR of Scott and Johnstone \cite{Scott2007} treats items $A \to \alpha \bullet  B \beta$ as intermediary nonterminals and performs path reductions in steps of 2, guaranteeing an at worst cubic runtime.

\subsection{The Boolean LR automaton}

We build our solution for Boolean grammars on the foundations laid by the $\epsilon$-GLR and the BRNGLR, namely
\begin{enumerate}
    \item never create an edge in the GSS for nulled inputs and
    \item never perform reductions of length greater than 2.
\end{enumerate}

Nullability of symbols is a property of the grammar and not the input, therefore it is possible to precompute this knowledge, for example by explicit evaluation of the $\Phi$ operator on logic rules given at the end of the previous part for the empty string, repeated until the interpretation has converged (i.e. does not change between successive evaluations; this must happen in at most $\mathbf{V}$ steps). Notice that $\epsilon$ is always positively nullable, terminals are never so.

Given the generalized structure of a Boolean grammar in the sense that we allow arbitrary formulas on the right-hand side, the items that form a state of the automaton will be labeled with arbitrary expressions that appear on the right-hand sides of grammar rules. As in our three-valued setting the lack of a proof for truth is insufficient to derive falsity (which is different from not-truth), we also augment items with a \emph{sign} that signals whether derivations of the item should result in a positive or negative proof of the formula. An item is, therefore a triple consisting of a sign (either $+$ or $-$) indicating whether a positive or negative proof is expected; a grammar formula $\phi$ and position of the ``dot'', an index that  ranges from $0$ to $n$ (inclusive) for $n$-ary concatenations and one of $0$ or $1$ for other items, signaling how much of a given expression has been recognized.

The successor of an item is the item with the same sign and formula, and a dot that is one position ahead. A completion item is one where the dot has the highest possible index; it does not have a successor.

Similarly to the context-free case, which only has positive concatenation and disjunction (in the form of nondeterminism induced by transitions on multiple possible rules), the states (i.e. sets of items) are the closure of some initial ``seed'' items over rules that will be given shortly, and may be computed using iterated saturation. The items originally present in the state are referred to as kernel items, while those added via the closure are the derived items.

For items with non-trivial grammar formulas further derived items must be present in the state, and we will say they are generated by the item(s) that caused forced their inclusion. The parents of an item are the non-concatenation items that generate it. Intuitively, these items represent the transitive closure of the grammar expressions that may be required for the proof the kernel items.

We now go over the various expression types to detail how their child items are generated:
\begin{itemize}
    \item $\pm t$ where $t \in \Sigma \cup \{ \epsilon \}$: Terminal items serve as the trivial cases of the matching algorithm and generate no further items. An item $+\epsilon$ is never matched against the input as it is required nullable (positively matches only the empty string and negatively matches everything else); $-\epsilon$ matches any input segment that is not empty. The terminals match the respective single character in the input and negatively match everything else.
    \item $\pm X$ where $X \in \mathbf{V}$: In order to match a grammar variable, the expression on the right-hand side of its defining rule must be matched, therefore a grammar variable generates exactly one item, $\pm \phi$, where $\phi$ is the grammar expression on the right-hand side of the rule $X \to \phi$. Because of the well-formedness criterion on our definition of a Boolean grammar, there is exactly one such rule.
    \item $\pm \neg \phi$: A negated expression matches if and only if $\phi$ matches with the opposite sign, therefore negated items generate the expression without the negation but the opposite sign, i.e. $\mp \phi$.
    \item $\pm\bigvee_{i\leq n}\phi_i$ and $\pm\bigwedge_{i\leq n}\phi_i$: For these items to match, some or all of their formulas need to be satisfied over some string. While the reduction phase (how the results are aggregated) is different for these items, for the purpose of building the automaton, they are handled equivalently and generate all their subformulas without a change of sign.
\end{itemize}

Concatenations also generate items within the state as part of the closure, but the concatenation is not considered as a parent of the generated item. This is because the reducer handles concatenations differently from other kinds of formulae.
\begin{itemize}
    \item $+ \phi_1 \cdots \phi_n$: This is the classical case of concatenation whose subformulas must be matched sequentially. In the spirit of the $\epsilon$-GLR described in \cite{Nederhof1996}, whenever the language defined by the dotted subformula includes the empty string, the successor item, i.e. the item with the dot at the successive index, is also included in the closure.
    \item $- \phi_1 \cdots \phi_n$: A negative concatenation is proven over some string if and only if we are able to ascertain that in every possible partition of the string there is a substring that is excluded by the respective language. (We note, without proof, that negative concatenation is also associative.) Suppose that the state contains an item $- \phi_1 \cdots \phi_n$ with the dot before some subformula $\phi_r$. To obtain a negative proof, either a negative proof of $\phi_r$ must be obtained, or the proof of $\phi_r$ may be skipped entirely and only the remainder of the rule (that is $\phi_{r + 1} \cdots \phi_n$) be matched. Given that we need to consider skipping input segments of zero length, we also unconditionally add the successor item to the current state under the assumption that it follows a zero-length skip.
\end{itemize}

The transitions from a state $s$ of the Boolean LR automaton are implied by the items of the state. For any item $\pm \bullet \phi$ that is not a concatenation, the automaton has a transition on $\pm\phi$ to a state with $\pm\phi \, \bullet $. For concatenation items, if the item is labeled $\pm \phi_1 \cdots \bullet  \phi_r \cdots \phi_n$, then there is a transition $\pm \phi_r$ to a state with the item $\pm \phi_1 \cdots \phi_r \bullet  \cdots \phi_n$. If the concatenation is negative, the transition is marked as optional: the transition may be taken over an arbitrary nonempty string in the input. Whenever this happens, the resulting edge in the GSS is marked with the special symbol $\ast$ and not the formula.

The rationale behind the optional (``don't care'') transitions is that in a negative concatenation it is always enough to negatively prove one subformula for a partition; all others may be mapped to arbitrary input segments. The marker $\ast$ will be referred to as a wildcard and a match marked with $\ast$ a wildcard match.

The pseudocode for building the automaton is presented in \textbf{Algorithm \ref{alg:automaton}}.

\begin{algorithm}[htpb]
\caption{Construction of the Boolean LR automaton.}
\label{alg:automaton}
%\removealgoends
\begin{algorithmic}
    \Procedure{Build-Automaton}{$S$}
        \State{\textbf{let} $s\sb{0}$ be a state with an empty kernel and items $+\bullet S$ and $-\bullet S$}
        \While{there is an unprocessed state $s$}
            \State \Call{Closure}{$s$}
        \EndWhile
    \EndProcedure
\end{algorithmic}
\begin{algorithmic}
    \Procedure{Closure}{$s$}
        \While{there is an unprocessed item $\iota$ in $s$ that is not a completion item}
            \If{$\iota$ is labeled $-\epsilon$ or $\pm c$ where $c \in \Sigma$}
                \State{\textbf{add} the completion of $\iota$ to the transition on the label of $\iota$}
            \ElsIf{$\iota$ is labeled $\pm A$ where $A \in \mathbf{V}$}
                \State{\textbf{create} a new item $\iota'$ in $s$ labeled $\pm \phi$ where there is a rule $A \to \phi$ in the grammar}
                \State{\textbf{add} the completion of $\iota$ to the transition on $\pm A$}
            \ElsIf{$\iota$ is labeled $\pm \neg \phi$}
                \State{\textbf{add} a new item $\iota'$ labeled $\mp\phi$ in $s$}
                \State{\textbf{add} the completion of $\iota$ to the transition on $\pm \neg \phi$}
            \ElsIf{$\iota$ is labeled $\pm \phi\sb{1} \lor \cdots \lor \phi\sb{n}$ or $\pm \phi\sb{1} \land \cdots \land \phi\sb{n}$}
                \For{$\phi$ \textbf{in} $\phi\sb{1}, \ldots, \phi\sb{n}$}
                    \State{\textbf{add} a new item $\iota'$ labeled $\pm\phi$ in $s$}
                \EndFor
                \State{\textbf{add} the completion of $\iota$ to the transition on the label of $\iota$}
            \ElsIf{$\iota$ is labeled $+ \phi\sb{1} \cdots \bullet  \phi\sb{r} \cdots \phi\sb{n}$}
                \State{\textbf{add} a new item labeled $+\phi\sb{r}$ to $s$}
                \State{\textbf{add} the successor of $\iota$ to the transition on $+\phi\sb{r}$}
                \If{$\phi\sb{r}$ is nullable \textbf{and} ($r < n$ \textbf{or} $\exists$ a right-nullable kernel item $+ \phi\sb{1} \cdots \phi\sb{n}$ in $s$)}
                    \State{\textbf{add} the successor of $\iota$ to $s$}
                \EndIf
            \ElsIf{$\iota$ is labeled $- \phi\sb{1} \cdots \bullet  \phi\sb{r} \cdots \phi\sb{n}$}
                \State{\textbf{add} a new item labeled $-\phi\sb{r}$ \textbf{to} $s$}
                \State{\textbf{add} successor of $\iota$ \textbf{to} the transition on $-\phi\sb{r}$}
                \State{\textbf{mark} the transition on $-\phi\sb{r}$ as optional}
                \If{$r < n$ \textbf{or} there is a kernel item $- \phi\sb{1} \cdots \phi\sb{n}$ in $s$}
                    \State{\textbf{add} the successor of $\iota$ to $s$}
                \EndIf
            \EndIf
        \EndWhile
    \EndProcedure
\end{algorithmic}
\end{algorithm}

\subsection{The Boolean GLR parser}

We now turn our attention to the actual parser in \textbf{Algorithm \ref{alg:parser}}. A match in a given input stream is identified by its left and right extents, which are the positions where the match begins and ends. As for any given left extent $i$ a trivial negative match may have almost any right extent $j \geq i$, the scanner part of the algorithm, even though progressing through the input in a left-to-right manner, finds for the current position $j$ all possible left extents $i < j$ where a trivial match may have begun. A trivial match here is either a positive or a negative match on $\epsilon$, a terminal symbol or a wildcard match.

Whenever an edge is created in the GSS, it signals the acquisition of new knowledge in the parsing process. It is necessary that further applications of this knowledge are investigated and the process is continued until no further derivations can be made. This is the job of the reducer.

Given that no reductions are performed over an interval of length zero, as these are precomputed, every reduction must involve at least one edge of the GSS. Whenever a new edge is created, the possible reductions starting with that edge are investigated. The set $\Delta_j$ at this point contains $\langle u, e \rangle$ pairs where $u$ is a vertex in the GSS and $e$ is an outgoing edge of $u$, pointing backwards, against the input direction. Such a pair is created exactly once for each edge in the GSS and serves as a work item for the reducer.

No constructs other than concatenation require more than one edge in the GSS to be traversed sequentially. As a first step, the reducer calls $\textsc{Finish-Reduction}$ with a formula $\pm\phi$ to sort out every reduction that is not a concatenation. The purpose of $\textsc{Finish-Reduction}$ is to take a recognized formula and apply it to its parents that are not concatenations. The parent formulas to be substituted into should be precomputed, but even searching for them is constant time in the length of the input.

A variable is considered matched whenever its definition is matched, therefore if one of the parents is a variable, an edge is immediately created that represents the match. Negation is similarly simple, upon matching $\pm\phi$ a transition on $\mp \neg\phi$ is recorded in the GSS. If the parent is a positive disjunction or a negative conjunction, matching the child immediately causes an edge to be created. These are collectively referred to as existential reductions.

Suppose that the parent is either a negative disjunction or a positive conjunction. These items require that all children are matched before the parent edge is created. (Note that child matches may end at different nodes in the GSS; this is no problem as long as these nodes belong to the same generation, and therefore cover the same part of the input.) Therefore, when processing these reductions, we only record in $u$ that one of the subformulas was matched, and only create the parent edge when records for all subformulas are present. These records are invalidated whenever the parser position advances, as matches on a segment $(i, j)$ are not meaningful for any other $(i, j')$.

So far we have discussed how reductions for the non-concatenation items are performed. We will now switch our attention to concatenations, as path tracing is not handled by the $\textsc{Finish-Reduction}$ function. For any formula other than negation, an edge will only satisfy an item of the same sign as the edge's label. Whenever an edge $e$ from $u$ to $v$ is created, where $u$ is a node in the current generation, only a reduction via the respective sign needs to be considered.

As the edge $e$ represents a transition of the underlying automaton, if $e$ is labeled $+\phi_r$, there must be some item $+\cdots \phi_r \bullet  \cdots$ in the kernel of $u$. The edge $e$ may be the last edge of such a reduction only if the rule is right-nullable, i.e. all of $\phi_{r + 1} \cdots \phi_n$ are positively nullable. In this case we shall traverse the edge and call the function $\textsc{Extend-Positive-Reduction}$, which, if the concatenation is fully reduced (i.e. $r = 1$) allows parent items to progress by invoking $\textsc{Finish-Reduction}$, otherwise it merely queues the rest of the rule for further progressing. Notice that even though we use the same set $\Delta_j$ for the queue as $\textsc{Create-Edge}$, this causes no confusion, as these items are of the shape $\langle v, \pm\phi\sb{1} \cdots \bullet  \phi\sb{r} \cdots \phi\sb{n} \rangle$, i.e. they do not name the specific edge the reduction should be continued on.

The function $\textsc{Continue-Positive-Reduction}$ takes a partially completed reduction that already has had at least one edge matched and traces the path further either by matching edges in the GSS or by eliminating them if they're positively nullable.

The last piece of the puzzle is the negative deduction of a concatenation. In order to prove for some segment $(i, j)$ that a concatenation does not hold, one must prove that in every partition of that segment there is at least one part that negatively matches. The problem with the naïve approach of simply enumerating all partitions is that there is $O(|w|^{n - 1})$ many of them, where $n$ is the number of concatenated entities. That's way too many. Luckily, the binarization technique is also applicable for negative concatenation, as the operation, like the positive case, remains associative.

Suppose that $-\phi_1 \cdots \phi_n$ is to be proven over some segment ($i, j$). It is clear that whatever partition one chooses, it has a position, call it $k$ that splits the interval into subintervals $(i, k)$ and $(k, j)$ such that either $-\phi_1$ matches over $(i, k)$ or $-\phi_2 \cdots \phi_n$ does over $(k, j)$. If one is able to prove $-\phi_2 \cdots \phi_n$ over $(k, j)$, then for any choice of $i \leq k$ $-\phi_1 \cdots \phi_n$ holds over $(i, j)$ for that specific $k$. If the proof of $-\phi_2 \cdots \phi_n$ over $(k, j)$ was unsuccessful, then we are limited to choices of $i$ where $(i, j)$ matches $-\phi_1$. We call these suffix and prefix proofs, respectively, of the partitioning point $k$.

One has to do this for all $i \leq k \leq j$ to consider a concatenation negatively proven. Note that for a positive concatenation/negative disjunction the number of subproofs required to be reducibe depends on the number of subformulas (i.e. a proof of $+ \phi_1 \wedge \cdots \wedge \phi_n$ requires proofs of $+ \phi_1, \ldots, + \phi_n$ each, for a negative concatenation the number of subproofs is dependent on the length of the string it is being proven over ($j - i + 1$ for a segment with extents $i$ and $j$).

We note that as the parser progresses in the input, suffix proofs get invalidated as the $j$ in $(k, j)$ changes, but prefix proofs may be considered permanent.

\begin{algorithm}
\caption{The Boolean GLR parser}
\label{alg:parser}
%\removealgoends
\begin{algorithmic}
	\Procedure{Parse}{$S$}
		\State \Call{Build-Automaton}{$S$}
		\State{\textbf{create} a node labeled $s\sb{0}$ in $U\sb{0}$}
		\If{$\pm \epsilon$ matches $S$}
			\State {\textbf{yield} the sign of $\epsilon$ at position $0$}
		\EndIf
		\For{$j$ \textbf{in} $[1 .. |w|]$}
			\State{\Call{Shifter}{}}
			\State{\Call{Reducer}{}}
		\EndFor
	\EndProcedure
\end{algorithmic}

\begin{algorithmic}
	\Procedure{Shifter}{}
		\For {\textbf{each} node $u$ \textbf{in} generations $U\sb{i}$ \textbf{where} $i < j$}
			\For {\textbf{each} terminal transition $t$ from $u$}
				\If{$w[i,j]$ matches $t$}
					\State{\Call{Create-Edge}{$u, t$}}
				\EndIf
			\EndFor
			\For {\textbf{each} optional transition from $u$}
				\State{\Call{Create-Edge}{$u, \ast$}}
			\EndFor
		\EndFor
	\EndProcedure
\end{algorithmic}
\begin{algorithmic}
    \Procedure{Create-Edge}{$u, l$}
		\State{\textbf{let} $v$ be the node in $U\kern-1pt\sb{j}$ reached by the transition on $l$ from $u$}
		\State{\textbf{add} an edge $e$ labeled $l$ from $v$ to $u$}
		\State{\textbf{add} $\langle v,e \rangle$ \textbf{to} $\Delta_j$}
	\EndProcedure
\end{algorithmic}

\begin{algorithmic}
    \Procedure{Reducer}{}
		\While{there is a pending reduction $\langle u, e \rangle$ in $\Delta_j$}
			\State{\Call{Finish-Reduction}{$e.target, e.label$}}
			\If{$e$ is positive}
				\For {$\iota$ \textbf{in} $u.kernel$ labeled $+\phi\sb{1} \cdots \phi\sb{r} \bullet  \cdots \phi\sb{n}$ \textbf{and} $+\epsilon$ matches $\phi\sb{r+1} \cdots \phi\sb{n}$}
					\State{$\Call{Extend-Positive-Reduction}{e.target, +\phi\sb{1} \cdots \bullet  \phi\sb{r} \cdots \phi\sb{n}}$}
				\EndFor
				\While{there is an unprocessed continuation $\langle v, +\phi\sb{1} \cdots \phi\sb{r} \bullet  \cdots \phi\sb{n} \rangle$ in $\Delta_j$}
					\State{\Call{Continue-Positive-Reduction}{$v, +\phi\sb{1} \cdots \phi\sb{r} \bullet  \cdots \phi\sb{n}$}}
				\EndWhile
			\Else
				\For {$\iota$ \textbf{in} $u.kernel$ labeled $-\phi\sb{1} \cdots \phi\sb{r} \bullet  \cdots \phi\sb{n}$}
					\If{$e$ is labeled $-\phi\sb{r}$ \textbf{or} $e$ is labeled $\ast$ \textbf{and} $-\epsilon$ matches $\phi\sb{r+1} \cdots \phi\sb{n}$}
						\If{$e$ is labeled $-\phi\sb{r}$}
							\State{\textbf{permanently mark} position $j$ as complete for $-\phi\sb{1} \cdots \bullet  \phi\sb{r} \cdots \phi\sb{n}$ at $v$}
						\EndIf
						\State{\Call{Extend-Negative-Reduction}{$e.target, -\phi\sb{1} \cdots \bullet  \phi\sb{r} \cdots \phi\sb{n}, j$}}
					\EndIf
				\EndFor
				\While{there is an unprocessed continuation $\langle v, -\phi\sb{1} \cdots \phi\sb{r} \bullet  \cdots \phi\sb{n} \rangle$ in $\Delta_j$}
					\State{\Call{Continue-Negative-Reduction}{$v, -\phi\sb{1} \cdots \phi\sb{r} \bullet  \cdots \phi\sb{n}$}}
				\EndWhile
			\EndIf
		\EndWhile
    \EndProcedure
\end{algorithmic}
\end{algorithm}

\setcounter{algorithm}{1}
\begin{algorithm}
\caption{The Boolean GLR parser (cont.)}
%\removealgoends

\begin{algorithmic}
    \Procedure{Finish-Reduction}{$u, \phi$}
		\State{\textbf{let} $\iota$ be the item in $u$ that is labeled $\bullet \phi$}
		\For{$\iota'$ \textbf{in} $\iota.parents$}
			\If{$\iota'$ is a variable, negation, positive disjunction or negative conjunction}
				\State{\Call{Create-Edge}{$u, \phi$}}
				\If{$\phi = \pm S$}
					\State{\textbf{yield} the sign of $\phi$ at position $j$}
				\EndIf
			\Else
				\State{\textbf{mark} the subformla $\phi$ as complete for $\iota'$ in $u$}
				\If{all subformulas of $\iota'$ are complete in $u$}
					\State{\Call{Create-Edge}{$u, \phi$}}
				\EndIf
			\EndIf
		\EndFor
	\EndProcedure
\end{algorithmic}
\begin{algorithmic}
	\Procedure{Continue-Positive-Reduction}{$v, +\phi\sb{1} \cdots  \phi\sb{r} \bullet  \cdots \phi\sb{n}$}
		\If{there is an item $+\phi\sb{1} \cdots \bullet  \phi\sb{r} \cdots \phi\sb{n}$ in $v$ \textbf{and} $+\epsilon$ matches $\phi\sb{r}$}
			\State{\Call{Extend-Positive-Reduction}{$v, +\phi\sb{1} \cdots \bullet  \phi\sb{r} \cdots \phi\sb{n}$}}
		\EndIf
		\If{$+\phi\sb{1} \cdots \phi\sb{r} \bullet  \cdots \phi\sb{n}$ is a kernel item in $v$}
			\For{$e$ \textbf{in} $v.edges$}
				\State{\Call{Extend-Positive-Reduction}{$e.target, +\phi\sb{1} \cdots \bullet  \phi\sb{r} \cdots \phi\sb{n}$}}
			\EndFor
		\EndIf
	\EndProcedure
\end{algorithmic}
\begin{algorithmic}
    \Procedure{Continue-Negative-Reduction}{$v, -\phi\sb{1} \cdots  \phi\sb{r} \bullet  \cdots \phi\sb{n}$}
		\State{\textbf{let} $i$ $\gets$ the position of $v$}
		\If{$v$ contains the item $-\phi\sb{1} \cdots \bullet  \phi\sb{r} \cdots \phi\sb{n}$}
			\State{\Call{Extend-Negative-Reduction}{$v, -\phi\sb{1} \cdots \bullet  \phi\sb{r} \cdots \phi\sb{n}, i$}}
		\EndIf
		\If{$v$ contains the kernel item $-\phi\sb{1} \cdots \phi\sb{r} \bullet  \cdots \phi\sb{n}$}
			\For{$e$ \textbf{in} $v.edges$ \textbf{where} $e$ is labeled $-\phi\sb{r}$ \textbf{or} $e$ is labeled $\ast$}
				\State{\Call{Extend-Negative-Reduction}{$e.target, -\phi\sb{1} \cdots \bullet  \phi\sb{r} \cdots \phi\sb{n}, i$}}
			\EndFor
		\EndIf
	\EndProcedure
\end{algorithmic}
\begin{algorithmic}
    \Procedure{Extend-Positive-Reduction}{$v, +\phi\sb{1} \cdots \bullet  \phi\sb{r} \cdots \phi\sb{n}$}
		\If{$r = 1$}
			\State{\Call{Finish-Reduction}{$v, +\phi\sb{1} \cdots \phi\sb{n}$}}
		\Else
			\State{\textbf{add} $\langle v, \pm\phi\sb{1} \cdots \bullet  \phi\sb{r} \cdots \phi\sb{n} \rangle$ \textbf{to} $\Delta_j$}
		\EndIf
	\EndProcedure
\end{algorithmic}
\begin{algorithmic}
    \Procedure{Extend-Negative-Reduction}{$v, -\phi\sb{1} \cdots \bullet  \phi\sb{r} \cdots \phi\sb{n}, p$}
		\State{\textbf{mark} position $p$ as complete for $-\phi\sb{1} \cdots \bullet  \phi\sb{r} \cdots \phi\sb{n}$ at $v$}
		\If{$r = n$ \textbf{or} $-\phi\sb{1} \cdots \bullet  \phi\sb{r} \cdots \phi\sb{n}$ just got completed at $v$}
			\If{$r = 1$}
				\State{\Call{Finish-Reduction}{$v, -\phi\sb{1} \cdots \phi\sb{n}$}}
			\Else
				\State{\textbf{add} $\langle v, -\phi\sb{1} \cdots \bullet  \phi\sb{r} \cdots \phi\sb{n} \rangle$ \textbf{to} $\Delta_j$}
			\EndIf
		\EndIf
	\EndProcedure
\end{algorithmic}
\end{algorithm}

\subsection{Notes}

\subsubsection{Generating parse trees}

Our algorithm does not deal with the construction of parse trees. As the aptly titled paper \cite{Scott2010} states, ``[r]ecognition is not parsing'', and we do indeed refer to our algorithm as a parser, rather than a recognizer, whereas it is, in a strict sense, the latter. Our excuse for doing so is that through the application of the usual techniques, it should not pose a significant technical challenge to turn the algorithm into an actual parser; the algorithm is structured and the GSS is constructed in a way that contains all the information that would be included in a parse tree. We therefore consider the implementation of parse trees a technicality that was omitted for brevity, but should not be hard to implement, should the reader desire to.

\subsubsection{Optimization opportunities}

We mention two possible opportunities for the optimization of the algorithm.

The first one involves the building of the Boolean LR automaton. Suppose that, for example, a state $s$ contains an item $+ \phi_1 \cdots \bullet  (\psi_1 \vee \psi_2) \cdots \phi_n$. Then, by closure, it also contains $+ \bullet  \psi_1 \vee \psi_2$ and then $+ \bullet  \psi_1$ and $+ \bullet  \psi_2$. Suppose $\psi_1$ is matched. The automaton currently has a transition on $+\psi_1$ to a state $p$ with the item $+ \psi_1 \bullet $, where it is reduced, trivially traced back to $s$, it is found that it has an existentially reducible parent $+ \bullet  \psi_1 \vee \psi_2$, which causes another transition on the formula $+ \psi_1 \vee \psi_2$ to a state $q$ with items $+ \psi_1 \vee \psi_2 \bullet $ and $+ \phi_1 \cdots (\psi_1 \vee \psi_2) \bullet  \cdots \phi_n$.

Notice, however, that when $+\psi_1$ is matched, it is always the case that $+ \psi_1 \vee \psi_2$ is matched, too, therefore it is possible to transition on both into at the same time. A drawback of this optimization is that the GSS would lose some of its structure, making the potential recovery of a parse tree harder.

The second optimization involves the tracing of paths in the GSS during the reduction phase. As the GSS is an append-only structure whose new edges are drawn only to the current generation, which is monotonously moving to the right, given any pending concatenation $\langle v, \pm\phi\sb{1} \cdots \bullet  \phi\sb{r} \cdots \phi\sb{n} \rangle$ the set of GSS nodes where the paths reading $\phi_{r-1}, \ldots, \phi_1$ (in order of backwards traversal) may end does not change as the algorithm progresses. It would therefore be possible to build these sets progressively as part of $\textsc{Create-Edge}$, in a manner similar to \cite{aycock_practical_2002}. This space-time tradeoff drastically reduces the time spent searching for paths in the GSS at the cost of an extra $O(|w|^2)$ storage.

\subsubsection{Complexity bounds}

Any generation of the GSS may contain at most $Q$ nodes, where $Q$ is the number of states of the Boolean LR automaton. For a string of length $|w|$, the largest possible number of different edges is of order $O({|w|}^2 \cdot Q \cdot F)$ where $F$ is the cardinality of the set of all possible labels of edges. Note that both $Q$ and $F$ are constant for any given grammar, therefore the size of the GSS is of order $O({|w|}^2)$.

As for the time complexity, assume a given $j$. The $\textsc{Shifter}$ runs in time $O(j \cdot Q\cdot F)$ in worst case, with $\textsc{Create-Edge}$ being $O(1)$. In the reducer, $\textsc{Finish-Reduction}$ is once again of $O(1)$ complexity (the iteration on the parent items is invariant with respect to the input length) that is called for at most $O(j \cdot Q\cdot F)$ times for each edge pointing away from generation $j$. The amount of work done on the concatenations are bounded by the size of $\Delta_j$, which are at most of size $O(j)$. For each pending reduction $\langle v, \pm\phi\sb{1} \cdots \bullet  \phi\sb{r} \cdots \phi\sb{n} \rangle$ the only operation that is not of constant time is the loop on the outgoing edges of $v$ in the $\textsc{Continue}$ methods, meaning there is $O(j^{\kern1pt2})$ steps performed overall for each concatenation. Summing up all the above results in an $O({|w|}^3)$ runtime, as $j$ runs over each position.

\section{Summary}

Boolean grammars are a straightforward generalization of context-free grammars that both allow the description of some languages that are not context-free, and simplify the description of others that are. The introduction of negation, however opens up the possibility of contradictory grammars that have no classical solution. An approach based on three-valued logic, where grammar rules are taken as a system of logic equations, always produces a model through the iteration of a simple substitutive process. The words found as being included or excluded from the language are exactly those entailed by the logic equations. Containment in the three-valued sense is always decidable within tame polynomial bounds for any given word of the alphabet and therefore serves as a suitable basis for a parser.

We provided a short overview of the logical foundations of the three-valued interpretation of Boolean grammars and gave an efficient algorithm from the GLR family of constructions that is able to determine the containment status of a string within cubic polynomial bounds.

\bibliographystyle{eptcs}
\bibliography{boolglr}

\end{document}